\documentclass[conference, 10pt]{IEEEtran}

\usepackage[utf8]{inputenc} 
\usepackage[T1]{fontenc}
\usepackage{url}
\usepackage{ifthen}
\usepackage{cite}
\usepackage{flushend}
\usepackage[cmex10]{amsmath} 
\usepackage[usenames,dvipsnames]{xcolor}
\usepackage{amsmath,amsthm}
\usepackage{bm}
\usepackage{amssymb}
\usepackage{mathtools}
\usepackage{caption}
\usepackage{mathabx}
\usepackage{dsfont}
\usepackage{graphicx}
\usepackage{epstopdf}
\usepackage{tikz}
\usepackage{scalerel}
\usepackage[ruled,vlined]{algorithm2e}
\usepackage{subcaption}
\usepackage{float}
\restylefloat{table}
\usepackage{tabularx}

\newcommand{\sref}[1]{Section~\ref{#1}}

\newtheorem{theorem}{Theorem}

\newtheorem{lemma}{Lemma}

\newenvironment{sproof}{%
	\proof}{\endproof}

\IEEEoverridecommandlockouts
\usepackage[left=0.63in,right=0.63in,top=0.7in,bottom=0.7in]{geometry}

\allowdisplaybreaks
\begin{document}
	\title{Secure Distributed/Federated Learning: Prediction-Privacy Trade-Off for Multi-Agent System}
	\author{
		\IEEEauthorblockN{Mohamed Ridha Znaidi$^{\ast}$, Gaurav Gupta$^{\ast}$, and Paul Bogdan}
		\IEEEauthorblockA{Ming Hsieh Department of Electrical and Computer Engineering,\\ University of Southern California, Los Angeles, CA 90089, USA\\
			Email:\{znaidi, ggaurav, pbogdan\}@usc.edu}
		\thanks{$^{\ast}$ authors contributed equally.}
	}  
	\maketitle
	
	\begin{abstract}
		Decentralized learning is an efficient emerging paradigm for boosting the computing capability of multiple bounded computing agents. In the big data era, performing inference within the distributed and federated learning (DL and FL) frameworks, the central server needs to process a large amount of data while relying on various agents to perform multiple distributed training tasks. Considering the decentralized computing topology, privacy has become a first-class concern. Moreover, assuming limited information processing capability for the agents calls for a sophisticated \textit{privacy-preserving decentralization} that ensures efficient computation. Towards this end, we study the \textit{privacy-aware server to multi-agent assignment} problem subject to information processing constraints associated with each agent, while maintaining the privacy and assuring learning informative messages received by agents about a global terminal through the distributed private federated learning (DPFL) approach. To find a decentralized scheme for a two-agent system, we formulate an optimization problem that balances privacy and accuracy, taking into account the quality of compression constraints associated with each agent. We propose an iterative converging algorithm by alternating over self-consistent equations. We also numerically evaluate the proposed solution to show the privacy-prediction trade-off and demonstrate the efficacy of the novel approach in ensuring privacy in DL and FL.
	\end{abstract}
	\IEEEpeerreviewmaketitle
	\bstctlcite{IEEEexample:BSTcontrol}
	\section{Introduction}
	\label{sec:intro}
	Understanding uncertain and highly-dynamic environments requires 
	processing in real-time a plethora of large volume of heterogeneous streaming data (e.g., audio, text, video) in a multi-agent fashion (i.e., relying on the collaboration among a team of autonomous agents). In such multi-agent scenarios, the computational complexity has evolved from simple binary object detection to more complex recognition of anomalous (e.g., malicious) activity, localization of mission-relevant entities, actors and targets, inference and prediction of their functions and intentions. 
	
	\begin{figure}
		\hspace{-0.95cm}
		\centering
		\includegraphics[width=0.95\linewidth, height = 1.75in]{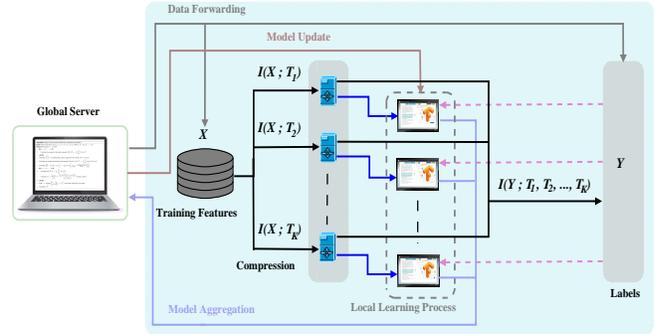}
		\vspace{0.4cm}
		\caption{The life-cycle of a secure DL/FL trained model involving $K$ multi-agent system aiming at making inference task. First, the global server forwards the training data to untrustworthy agents upon revealing limited information $\mathbf{T}_{1}$ to $\mathbf{T}_{K}$ and creating minimum inter-agent data sharing. Then, the agents perform local computation, the server aggregates all models to update the global model to be forwarded to all agents. The last two steps will be performed iteratively during the training process.}
		\label{fig:multiAgent}
	\end{figure}
	
	Recent research in distributed multi-agent computing led to introducing the Distributed and Federated Learning (DL and FL) as an emerging machine learning paradigms that use decentralization. On one hand, DL has been widely studied in several works \cite{Abadi,Shen,Bayen}. Many distributed optimization-based algorithms have been proposed to enhance the performance during the training process \cite{Zheng,Wu,ZhouZ,Alistarh}. Based on the topology algorithms, the decentralized learning algorithms can be classified into two categories (fixed topology algorithms \cite{Lian} and random topology algorithms \cite{Daily2018}).  
	On the other hand, FL involves training statistical models over decentralized devices \cite{mcmahan2017communication,konevcny2015federated,kairouz2019advances,li2020federated}. Contrary to conventional learning algorithms where the training model is computed using a central server, the FL framework proposes a collaborative training model where multiple clients are involved to train the model while the training data is decentralized. More specifically, a central server distributes training data into individual agents and then build a global model based on locally-computed model by every agent. 
	
	Along with the ongoing research in DL and FL, the reasoning about privacy in the real-world deployment is also rising. In the multi-agent scenario, the privacy concerns arise at the server level when the data center is bound to use computing agents that are untrustworthy. For example, government possessing private health records of citizens wish to distribute the data to computing agents such that they cannot recover the complete information. In another situation, a private firm intending to hire foreign computing centers does not wish to reveal the entire data, but carefully distribute it across various centers. In addition, they want to minimize the chances of foreign collaborations for reconstructing the various partly revealed data. Especially, in the case of an adversarial attack from an outsider, DL and FL have to ensure the privacy, 
	as a defense mechanism against attacks. The limited information transmission can be achieved by soft-clustering of the available data at the data-centers to introduce blurriness. However, the introduced compression (limiting information) often reduces the predictive quality of the data to make inference. Consequently, the problem of privacy-assurance in FL context while achieving good prediction has to be addressed carefully. 
	
	\subsection{Related work}
	As discussed above, decentralization mitigates some privacy guarantees resulting from discarding the conventional centralized learning assumption. This challenge has motivated many researchers to tackle the privacy problem in FL. Privacy assurance techniques widely used in machine learning are data anonymisation \cite{narayanan2008robust}, differential privacy (DP) \cite{dwork2006our}, secure multi-party computation (SMC) \cite{goldreich2009foundations}, and homomorphic encryption \cite{gentry2010computing}. In fact, the various proposed approaches to solve the privacy concerns in FL can be classified into two categories; \textit{global privacy} where the central server is trusted, and \textit{local privacy} where the central server might be malicious. Indeed, the privacy problem in FL has been addressed using the SMC protocol. In \cite{segal2017practical}, the authors introduce a protocol where the central server is not able to see any local updates, but can still observe the exact aggregated results at each round. Another solution is to use DP which is most widely used privacy approach in machine learning \cite{geyer2017differentially,thakkar2019differentially,bhowmick2018protection,li2019differentially,carlini2019secret,mcmahan2017learning,ghazi2019scalable}. The main advantage of DP is that it provides information theoretic guarantees about the data privacy. This technique helps in restraining the knowledge on whether a data sample is involved in the learning process or not. In \cite{agarwal2018cpsgd}, the authors introduced compression techniques added to DP for limiting the communication between multiples agents while performing stochastic gradient descent during the learning process. In this context, we address the trade-off between privacy and prediction in decentralized scenario via the distributed private FL (DPFL) formulation. More specifically, we propose a novel technique to be used in DL/FL to ensure privacy and to protect data effectively. The objective of the paper is to design an optimal scheme for forwarding the data (i.e., representations of relevant information) from a server to an untrustworthy multi-agent system under privacy constraints. This proposed approach is a key step that has to be done at the server level before performing DL/FL to preserve data privacy against any adversarial attacks. The privacy at the server level is achieved via (\textit{i}) transmitting limited information to the individual untrustworthy agents (compression), and (\textit{ii}) ensuring minimum inter-agent data sharing, relying on the principle that 
	``\textit{the fewer information sharing between the system's components, the higher its privacy}". The privacy is constrained by the requirement of collective information from the agents to recover relevant statistics for the inference task.
	
	\subsection{Motivation and Contribution}
	
	In the big data era, the DL and FL paradigms seem promising for improving learning from tremendous complex data streams. However, a critical design issue for decentralized architectures is privacy, especially when the data server does not wish to reveal the entire information to a computing agent. In this paper, we present an information-theoretic framework (i.e., DPFL) that investigates the prediction-privacy trade-off in DL and FL, in an untrustworthy multi-agent system. More specifically, our work deals with the model shown in Fig.\ref{fig:multiAgent}, where a server is trying to create a blurry representation of the data for $K$ untrustworthy agents. Each representation that will be sent to every agent has to (\textit{i}) satisfy a compression constraint, (\textit{ii}) minimize the inter-agent data sharing (i.e., we assume that different agents can share the same information up to a given level of privacy to inhibit harmful collaboration), and (\textit{iii}) carry only relevant information about the data $\mathbf{Y}$ of a given terminal. This framework can serve as a first step towards designing a more sophisticated algorithm applicable in cryptography. 
	Besides, our focus in this work is on a multi-agent system with $K=2$, although the case of a larger number of agents can be considered and studied similarly as described below by adding the appropriate constraints.
	
	The main contributions are; (\textit{i}) We present a new approach to tackle the problem of privacy in DL/FL, through the DPFL formulation. (\textit{ii}) For privacy,
	we perform soft-clustering of the server’s data for processing by
	the multi-agent system. (\textit{iii}) For the special case of a Gaussian distribution, we provide closed-form expressions for the iterative procedure.
	
	\section{Problem Formulation}
	\label{sec:probForm}
	In this section, we formulate the DPFL optimization problem for an architecture consisting of two-agents (i.e., $K=2$). We aim at finding the optimal stochastic mappings $\mathbf{T}_1$ and $\mathbf{T}_2$ (related to agents $1$ and $2$, respectively), that the server should perform. The aim of the server is to reveal limited information to each individual agents (to ensure privacy), as well as, ensure minimum data sharing between $\mathbf{T}_{1}$ and $\mathbf{T}_{2}$ to prevent evil collaboration for possible data breach (i.e., by minimizing the mutual information of $\mathbf{T}_{1}$ and $\mathbf{T}_{2}$). At the same time, the information should be well distributed in the multi-agent setup to ensure collective prediction accuracy for inference task.

	\subsection{Optimization Problem Setup}
	\label{parag:optimProb}
	
	Given that the agents are untrustworthy in the multi-agent scenario, our goal is to design two compact representations $\mathbf{T}_1$ and $\mathbf{T}_2$ from the data $\mathbf{X}$ to predict $\mathbf{Y}$. In what follows, since $\mathbf{T}_1$ and $\mathbf{T}_2$ are mappings of $\mathbf{X}$, we use the conditional independence conditions $\mathbf{T}_{1}\perp\mathcal{X}\backslash\{\mathbf{T}_{1}\}\vert\mathbf{X}$, $\mathbf{T}_{2}\perp\mathcal{X}\backslash\{\mathbf{T}_{2}\}\vert\mathbf{X}$ where $\mathcal{X}$ is the set of all random variables except $\mathbf{X}$. Similarly, since $\mathbf{X}$ has the original information for $\mathbf{Y}$, we take $\mathbf{Y}\perp\mathcal{X}\backslash\{\mathbf{Y}\}\vert\mathbf{X}$. 
	
	The $\mathbf{T}_1$ and $\mathbf{T}_2$ are such that limited information is revealed individually, which we achieve through bounding the information terms $I(\mathbf{X};\mathbf{T}_{1}), I(\mathbf{X};\mathbf{T}_{2})$ (i.e., guaranteeing a compression level at each agent). Additionally, to prevent malicious behavior from the collaboration of untrustworthy agents (also, to defend the entire system against an adversarial attack that may affect one agent and propagates progressively to remaining agents), we limit the correlation by forcing $I(\mathbf{T}_{1};\mathbf{T}_{2})$ to meet a certain requirement in order to create maximum independence across agents. Recall that the information sent to agents is used to perform the inference task. Hence, the server has to send relevant information to agents for learning purposes. Also, by limiting the amount of information received by each agent due to privacy enforcement, the prediction performed by the multi-agent setup may not be sufficient for the inference task. Consequently, the joint prediction defined as $I(\mathbf{Y};\mathbf{T}_{1},\mathbf{T}_{2})$ has to be maximized. Thus, the optimization problem for the DPFL can be formulated as follows:
	\begin{align}
		\begin{aligned}
			&\underset{{p(\mathbf{T}_{1}|\mathbf{X})},{p(\mathbf{T}_{2}|\mathbf{X})}}{\text{max}}
			& & I(\mathbf{Y};\mathbf{T}_{1},\mathbf{T}_{2})  \\
			& \text{~~subject to}
			&& I(\mathbf{X};\textbf{T}_{1}) & \leq r_1\\
			&&& I(\mathbf{X};\mathbf{T}_{2}) & \leq r_2\\
			&&& I(\mathbf{T}_{1};\mathbf{T}_{2})  & \leq \epsilon_{\hspace{5pt}}
		\end{aligned}
		\label{eqn:optimProb}
	\end{align}
	where $r_1$ and $r_2$ designate the compression level corresponding to agents $1$ and $2$, respectively. The parameter $\epsilon$ ensures a requirement on the information sharing between agents i.e., privacy level. In the next section, we provide the solution for the aforementioned optimization problem.
	
	\section{Problem Solution}
	\label{sec:probSol}
	
	The DPFL problem in (\ref{eqn:optimProb}) is solved by writing the Lagrangian functional involving information theoretic terms. The functional for this optimization problem, ignoring constant terms, is written as follows
	\begin{align}
		&\mathcal{F}[p(\mathbf{T}_{1}|\mathbf{X}), p({\bf T}_{2}\vert\mathbf{X})]=-I(\mathbf{Y};\mathbf{T}_{1},\mathbf{T}_{2}) +\beta I(\mathbf{T}_{1};\mathbf{X})\nonumber\\ 
		&\qquad\qquad\qquad{+}\lambda I(\mathbf{T}_{2};\mathbf{X})
		+ \gamma I(\mathbf{T}_{1};\mathbf{T}_{2}),
		\label{eqn:lagrangeFunctional}
	\end{align}
	where $\beta, \lambda$ and $\gamma \geq 0$ are the Lagrange multipliers corresponding to the constraints in (\ref{eqn:optimProb}). We first provide the solution structure for the general discrete case and then also argue to apply it for the Gaussian distribution of the variables. The solution for the discrete case is inspired from the following result.
	
	\begin{theorem}
		The optimal solution that minimizes the functional $\mathcal{F}$ in (\ref{eqn:lagrangeFunctional}) satisfies the following self-consistent equations:
		\begin{align}
			&p(\mathbf{T}_{1}|\mathbf{X})=\frac{p(\mathbf{T}_{1})}{Z_1}\times\exp\left\{\frac{\gamma}{\beta}D_{KL}[p(\mathbf{T}_{2}\vert\mathbf{X})\vert\vert p(\mathbf{T}_{2}\vert\mathbf{T}_{1})] \right.\nonumber\\
			&\qquad\qquad \left. - \frac{1}{\beta}\mathbb{E}_{\mathbf{T}_{2}\vert\mathbf{X}}D_{KL}[p(\mathbf{Y}\vert\mathbf{X})\vert\vert p(\mathbf{Y}\vert\mathbf{T}_{1},\mathbf{T}_{2})]\right\}, \label{eqn:thm1_1}\\
			&p(\mathbf{T}_{2}|\mathbf{X})=\frac{p(\mathbf{T}_{2})}{Z_2} \times \exp\left\{\frac{\gamma}{\lambda}D_{KL}[p(\mathbf{T}_{1}\vert\mathbf{X})\vert\vert p(\mathbf{T}_{1}\vert\mathbf{T}_{2})] \right.\nonumber\\
			&\qquad\qquad \left. - \frac{1}{\lambda}\mathbb{E}_{\mathbf{T}_{1}\vert\mathbf{X}}D_{KL}[p(\mathbf{Y}\vert\mathbf{X})\vert\vert p(\mathbf{Y}\vert\mathbf{T}_{1},\mathbf{T}_{2})]\right\},
			\label{eqn:thm1_2}
		\end{align}
		where $Z_1$ and $Z_2$ are normalizing partition functions and $D_{KL}[.]$ designates the Kullback–Leibler (KL) divergence.
		\label{thm:generalmultiAgent}
	\end{theorem}
	\begin{sproof}
		We provide a brief outline of the steps involved in deriving the equations (\ref{eqn:thm1_1})-(\ref{eqn:thm1_2}). The Lagrangian for solving optimization problem in (\ref{eqn:optimProb}) is written by appending probability simplex constraints in the functional (\ref{eqn:lagrangeFunctional}). Next, the derivatives of the information terms appearing in the functional in (\ref{eqn:lagrangeFunctional}) are written as follows.
		\begin{align}
			\frac{\delta I(\mathbf{T}_{1};\mathbf{T}_{2})}{\delta p(\mathbf{T}_{1}|\mathbf{X})} &= p(\mathbf{X})\{D_{KL}[p(\mathbf{T}_{2}\vert\mathbf{X}\vert\vert p(\mathbf{T}_{2}))] \nonumber\\
			&\quad{-} D_{KL}[p(\mathbf{T}_{2}\vert\mathbf{X})\vert\vert p(\mathbf{T}_{2}\vert\mathbf{T}_{1})]\}, \label{eqn:thm1_pf_1}\\
			\frac{\delta I(\mathbf{T}_{1};\mathbf{X})}{\delta p(\mathbf{T}_{1}|\mathbf{X})} &= p(\mathbf{X})\text{log}(\frac{p(\mathbf{T}_{1}\vert\mathbf{X})}{p(\mathbf{T}_{1})}),\label{eqn:thm1_pf_2}\\
			\frac{\delta I(\mathbf{Y};\mathbf{T}_{1}\mathbf{T}_{2})}{\delta p(\mathbf{T}_{1}|\mathbf{X})} &= p(\mathbf{X})\mathbb{E}_{\mathbf{T}_{2}\vert\mathbf{X}}[D_{KL}[p(\mathbf{Y}\vert\mathbf{X})\vert\vert p(\mathbf{Y}\vert\mathbf{T}_{2})]] \nonumber\\
			& {-} \mathbb{E}_{\mathbf{T}_{2}\vert\mathbf{X}}D_{KL}[p(\mathbf{Y}\vert\mathbf{X})\vert\vert p(\mathbf{Y}\vert\mathbf{T}_{1},\mathbf{T}_{2})].
			\label{eqn:thm1_pf_3}
		\end{align}
		The derivative for the information term $I(\mathbf{X};\mathbf{T}_{2})$ with respect to (w.r.t.) $p(\mathbf{T}_{1}\vert\mathbf{X})$ is zero. Using the  identities in (\ref{eqn:thm1_pf_1})-(\ref{eqn:thm1_pf_3}), and then upon equating the derivative of the Lagrangian with zero, we obtain the self-consistent equation (\ref{eqn:thm1_1}). By symmetry, we use the same procedure for $p(\mathbf{T}_{2}\vert\mathbf{X})$ and obtain equation (\ref{eqn:thm1_2}).
	\end{sproof}
	
	The computations of the required probabilities appearing in the optimization problem (\ref{eqn:optimProb}) are done by using the self-consistent equations in Theorem\,\ref{thm:generalmultiAgent}. An iterative procedure similar to Blahut-Arimoto algorithm \cite{arimoto, blahut} is derived using equations (\ref{eqn:thm1_1}), (\ref{eqn:thm1_2}), and the conditional independence conditions in Section\,\ref{parag:optimProb}. The iteration steps are provided in the Appendix\,\ref{appn:iterProc}. Next, we show that the proposed iterative procedure is convergent.
	
	\begin{lemma}
		The iterative procedure using the self-consistent equations (\ref{eqn:thm1_1}), (\ref{eqn:thm1_2}) is guaranteed to converge to some stationary point.
	\end{lemma}
	\begin{sproof}
		We show convergence of the iterations by proving; (\textit{i}) functional in (\ref{eqn:lagrangeFunctional}) is lower bounded, and (\textit{ii}) the self-consistent equations (\ref{eqn:SCeqn1})-(\ref{eqn:SCeqn5}) monotonically decrease the functional.
		
		To see the lower-bounded nature of functional, we observe that it is positive combination of mutual information terms, and $I(\mathbf{Y};\mathbf{T}_{1},\mathbf{T}_{2})$ can be written as
		\begin{align*}
			&-I(\mathbf{Y};\mathbf{T}_{1},\mathbf{T}_{2}) = -I(\mathbf{X};\mathbf{Y})\\
			&\qquad+\mathbb{E}_{\mathbf{X}}\mathbb{E}_{\mathbf{T}_{1}\vert\mathbf{X}}\mathbb{E}_{\mathbf{T}_{2}\vert\mathbf{X}}\,D_{KL}[p(\mathbf{Y}\vert\mathbf{X})\vert\vert p(\mathbf{Y}\vert\mathbf{T}_{1},\mathbf{T}_{2})].
		\end{align*}
		Since $I(\mathbf{X};\mathbf{Y})$ is constant and $D_{KL}[.]\geq 0$, therefore, the functional in (\ref{eqn:lagrangeFunctional}) is always lower-bounded for positive Lagrange multipliers $\beta, \lambda, \gamma$. The detailed proof of the monotonicity is skipped due to space limitation, and the idea is inspired from the works of \cite{arimoto, tishby2000information}.
		The value of functional at the $t$-th iteration is denoted as $\mathcal{F}^{(t)}$, and a subsidiary functional $\mathcal{C}(t,t)$, similar to \cite{arimoto}, but with the optimization of functional in (\ref{eqn:lagrangeFunctional}), can be defined such that the first argument is related to probability functions in (\ref{eqn:SCeqn1})-(\ref{eqn:SCeqn2}) and the second argument to (\ref{eqn:SCeqn3})-(\ref{eqn:SCeqn5}). It can then be showed that
		\begin{eqnarray}
			\mathcal{F}^{(t)} = \mathcal{C}(t,t) \geq \mathcal{C}(t+1,t) \geq \mathcal{C}(t+1,t+1) = \mathcal{F}^{(t+1)},
		\end{eqnarray}
		\noindent where we perform the iterations by using the equations (\ref{eqn:SCeqn1})-(\ref{eqn:SCeqn5}) in this order.
	\end{sproof}
	We also show the DPFL for the case of continuous distribution of Gaussian in the following section.
	
	\subsection{Gaussian case}
	\label{ssec:gaussian}
	The results obtained in \sref{sec:probSol} are applied to the Gaussian distribution, where we assume that the input variables $\mathbf{X}$ and $\mathbf{Y}$ are zero-mean and jointly multivariate Gaussian. 
	
	Inspired from prior works \cite{globersonTishbyTR2004, chechik2005information}, we define the two representations $\mathbf{T}_{1}$ and $\mathbf{T}_{2}$ that ensure the DPFL to be jointly Gaussian with input $\mathbf{X}$. Consecutively, they can be written as affine transformation of the input as follows:
	\begin{eqnarray}
		\mathbf{T}_{1} &=& \mathbf{A}\mathbf{X} + \bm{\zeta}_{1},\label{eqn:gaussianBottle1}\\
		\mathbf{T}_{2} &=& \mathbf{B}\mathbf{X} + \bm{\zeta}_{2},\label{eqn:gaussianBottle2}
	\end{eqnarray}
	\noindent where $\mathbf{A}, \mathbf{B}$ are input coupling matrices for $\mathbf{T}_{1}, \mathbf{T}_{2}$, respectively. The variables $\bm{\zeta_{1}}, \bm{\zeta_{2}}$ are mean-centered Gaussian with covariance matrix $\mathbf{\Sigma}_{\bm{\zeta_{1}}}, \mathbf{\Sigma}_{\bm{\zeta_{2}}}$, respectively. For the task of characterizing the DPFL in the Gaussian setup, we realize that it is sufficient to estimate the coupling and associated covariance matrices of the remainder terms. The following result utilizes the general result in Theorem\,\ref{thm:generalmultiAgent} to achieve this task.
	
	\begin{theorem}
		The two Gaussian representations that ensure the DPFL in (\ref{eqn:gaussianBottle1}), (\ref{eqn:gaussianBottle2}) for given Lagrangian parameters tuple $(\beta, \lambda, \gamma)$ is determined using the following iterative procedure at the $t$-th iteration.
		\begin{align}
			\mathbf{\Sigma}_{\bm{\zeta}_{1}}^{-1\,(t+1)} &= \mathbf{\Sigma}_{\mathbf{T}_{1}}^{-1\,(t)} - \frac{\gamma}{\beta}\mathbf{\Xi}_{1}^{T\,(t)}\mathbf{\Sigma}_{\mathbf{T}_{2}\vert\mathbf{T}_{1}}^{-1\,(t)}\mathbf{\Xi}_{1}^{(t)} \nonumber \\
			& \quad + \frac{1}{\beta}\mathbf{\Psi}_{1}^{T\,(t)}\mathbf{\Sigma}_{\mathbf{Y}\vert\mathbf{T}_{1},\mathbf{T}_{2}}^{-1\,(t)}\mathbf{\Psi}_{1}^{(t)},\label{eqn:thm2_1}\\
			\mathbf{A}^{(t+1)} &= \mathbf{\Sigma}_{\bm{\zeta}_{1}}^{(t+1)}[-\frac{\gamma}{\beta}\mathbf{\Xi}_{1}^{T\,(t)}\mathbf{\Sigma}_{\mathbf{T}_{2}\vert\mathbf{T}_{1}}^{-1\,(t)}\mathbf{B}^{(t)} \nonumber \\
			& \quad + \frac{1}{\beta}\mathbf{\Psi}_{1}^{T\,(t)}\mathbf{\Sigma}_{\mathbf{Y}\vert\mathbf{T}_{1},\mathbf{T}_{2}}^{-1\,(t)}(\mathbf{\Theta} - \bm{\Psi}_{2}^{(t)}\mathbf{B}^{(t)})],\label{eqn:thm2_2}\\
			\mathbf{\Sigma}_{\bm{\zeta}_{2}}^{-1\,(t+1)} &= \mathbf{\Sigma}_{\mathbf{T}_{2}}^{-1\,(t)} - \frac{\gamma}{\lambda}\mathbf{\Xi}_{2}^{T\,(t)}\mathbf{\Sigma}_{\mathbf{T}_{1}\vert\mathbf{T}_{2}}^{-1\,(t)}\mathbf{\Xi}_{2}^{(t)} \nonumber \\
			& \quad + \frac{1}{\lambda}\mathbf{\Psi}_{2}^{T\,(t)}\mathbf{\Sigma}_{\mathbf{Y}\vert\mathbf{T}_{2},\mathbf{T}_{1}}^{-1\,(t)}\mathbf{\Psi}_{2}^{(t)},\label{eqn:thm2_3}\\
			\mathbf{B}^{(t+1)} &= \mathbf{\Sigma}_{\bm{\zeta}_{2}}^{(t+1)}[-\frac{\gamma}{\lambda}\mathbf{\Xi}_{2}^{T\,(t)}\mathbf{\Sigma}_{\mathbf{T}_{1}\vert\mathbf{T}_{2}}^{-1\,(t)}\mathbf{A}^{(t+1)} \nonumber \\
			& \quad + \frac{1}{\lambda}\mathbf{\Psi}_{2}^{T\,(t)}\mathbf{\Sigma}_{\mathbf{Y}\vert\mathbf{T}_{1},\mathbf{T}_{2}}^{-1\,(t)}(\mathbf{\Theta} - \bm{\Psi}_{1}^{(t)}\mathbf{A}^{(t+1)})].
			\label{eqn:thm2_4}
		\end{align}
		\label{thm:gaussianBottle}
	\end{theorem}
	\begin{sproof}
		The Gaussian representations in (\ref{eqn:gaussianBottle1}), (\ref{eqn:gaussianBottle2}) are determined using Theorem\,\ref{thm:generalmultiAgent} and the definition of KL-divergence for Gaussian RVs.
		
		We write the following for the relevant KL-divergence terms in the Theorem\,\ref{thm:generalmultiAgent}.
		\begin{align}
			&D_{KL}[p(\mathbf{T}_{2}\vert\mathbf{X})\vert\vert p(\mathbf{T}_{2}\vert\mathbf{T}_{1})] = c_{1}(\mathbf{X}) \\
			&\hspace*{50pt}+ \frac{1}{2}\mathbf{T}_{1}^{T}\mathbf{\Xi}_{1}^{T}\mathbf{\Sigma}_{\mathbf{T}_{2}\vert\mathbf{T}_{1}}^{-1}\mathbf{\Xi}_{1}\mathbf{T}_{1} \nonumber \\
			& \hspace*{50pt} - \mathbf{T}_{1}^{T}\bm{\zeta}_{1}^{T}\mathbf{\Sigma}_{\mathbf{T}_{2}\vert\mathbf{T}_{1}}^{-1}\mathbf{B}\mathbf{X}, \label{eqn:KLDDefGauss1}\\
			&\mathbb{E}_{\mathbf{T}_{2}\vert\mathbf{X}}D_{KL}[p(\mathbf{Y}\vert\mathbf{X})\vert\vert p(\mathbf{Y}\vert\mathbf{T}_{1},\mathbf{T}_{2})] = c_{2}(\mathbf{X}) \nonumber \\
			&\quad +\frac{1}{2}\mathbf{T}_{1}^{T}\mathbf{\Psi}_{1}^{T}\mathbf{\Sigma}_{\mathbf{Y}\vert\mathbf{T}_{1}, \mathbf{T}_{2}}^{-1}\mathbf{\Psi}_{1}\mathbf{T}_{1} - \mathbf{T}_{1}^{T}\bm{\Psi}_{1}^{T}\mathbf{\Sigma}_{\mathbf{Y}\vert\mathbf{T}_{1}, \mathbf{T}_{2}}^{-1}\bm{\Theta}\mathbf{X} \nonumber\\
			&\quad + \mathbf{T}_{1}^{T}\bm{\Psi}_{1}^{T}\mathbf{\Sigma}_{\mathbf{Y}\vert\mathbf{T}_{1}, \mathbf{T}_{2}}^{-1}\bm{\Psi}_{2}\mathbf{B}\mathbf{X},
			\label{eqn:KLDDefGauss2}
		\end{align}
		\noindent where the definitions of the parameters are provided in the full version of the paper Appendix\,\ref{appn:paramDefGaussian}. 
		Upon substituting equations (\ref{eqn:KLDDefGauss1}), (\ref{eqn:KLDDefGauss2}) in (\ref{eqn:thm1_1}) (or iterative equation (\ref{eqn:SCeqn1})) and equating both sides, we obtain the iterative procedure of (\ref{eqn:thm2_1}) and (\ref{eqn:thm2_2}). By symmetry, we obtain the procedure for the mapping $\mathbf{T}_{2}$ as well.
	\end{sproof}
	With the developed theoretical solutions in Section\,\ref{sec:probSol}, we proceed forward to numerical results in the following section.
	\section{Simulations}
	\label{sec:sim}
	The solution for DPFL is simulated for the Gaussian distribution case. We take the input variables as $\mathbf{X}\in\mathbb{R^{N_{\mathbf{X}}}}$ and $\mathbf{Y}\in\mathbb{R^{N_{\mathbf{Y}}}}$, with $\mathbf{X}$ and $\mathbf{Y}$ being zero-mean jointly Gaussian. The input covariance matrices of appropriate dimensions are designed numerically.
	
	\begin{figure*}[ht]
		\centering
		\begin{subfigure}{.45\linewidth}
			\centering
			\includegraphics[width=\linewidth, height = 2in]{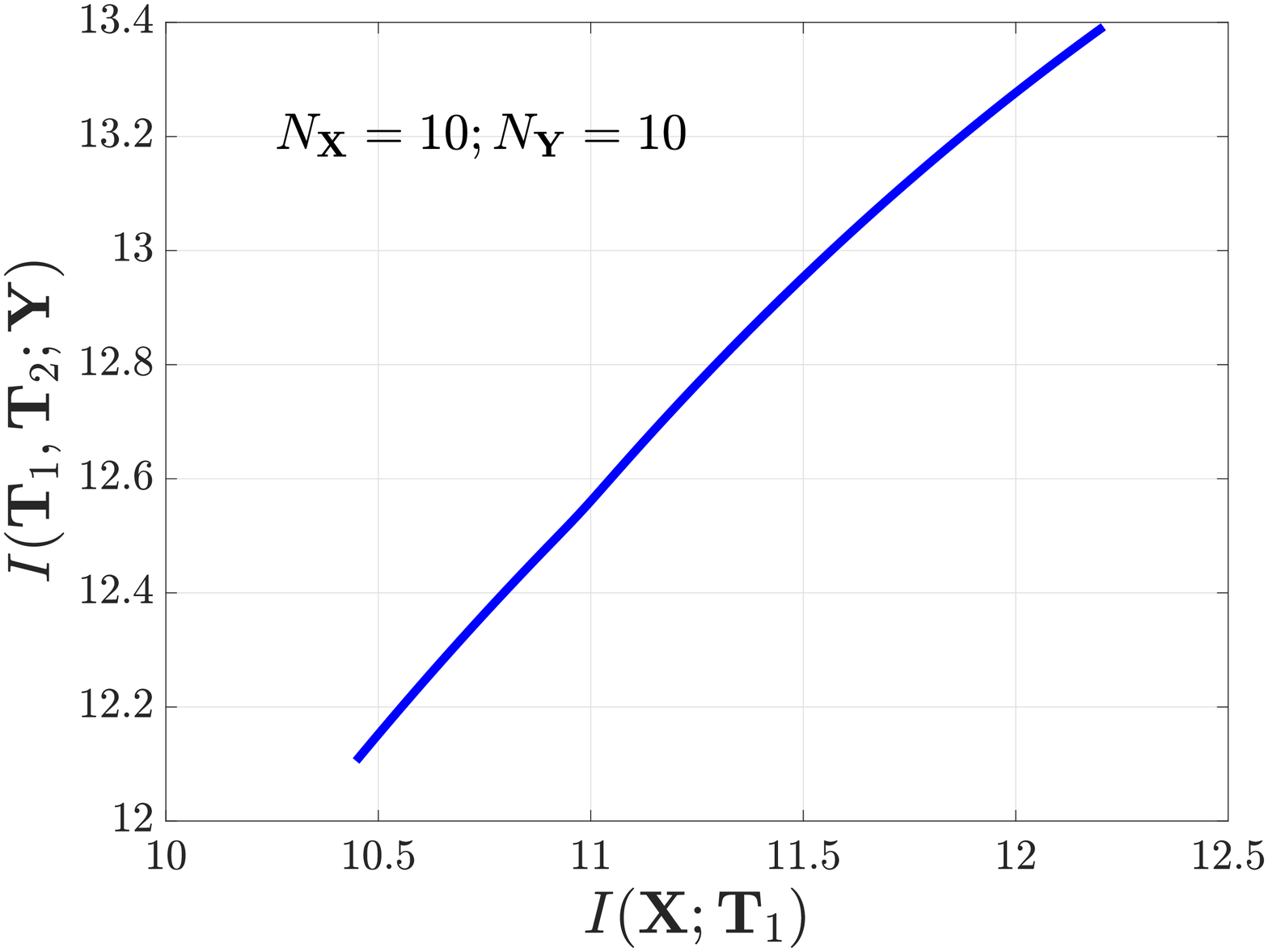}
			\caption{}
			\label{fig:plot1}
		\end{subfigure}%
		\hspace{20pt}
		\begin{subfigure}{.45\linewidth}
			\centering
			\includegraphics[width=\linewidth, height = 2in]{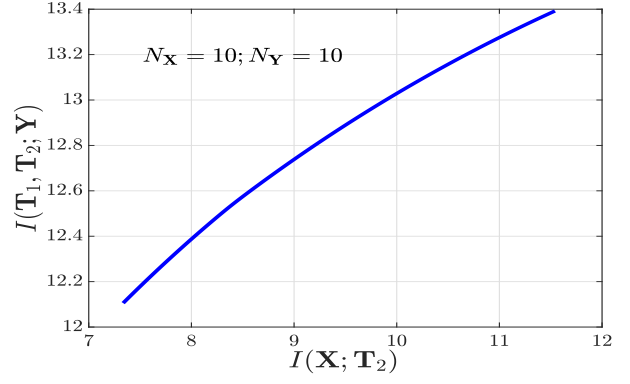}
			\caption{}
			\label{fig:plot2}
		\end{subfigure}\\
		\begin{subfigure}{.45\linewidth}
			\centering
			\includegraphics[width=\linewidth, height = 2in]{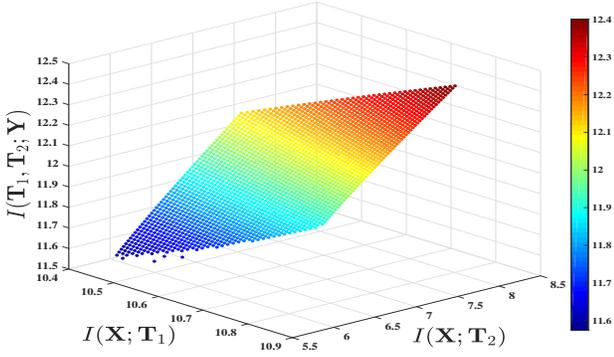}
			\caption{}
			\label{fig:plot3}
		\end{subfigure}
		\hspace{20pt}
		\begin{subfigure}{.45\linewidth}
			\centering
			\includegraphics[width=\linewidth, height = 2in]{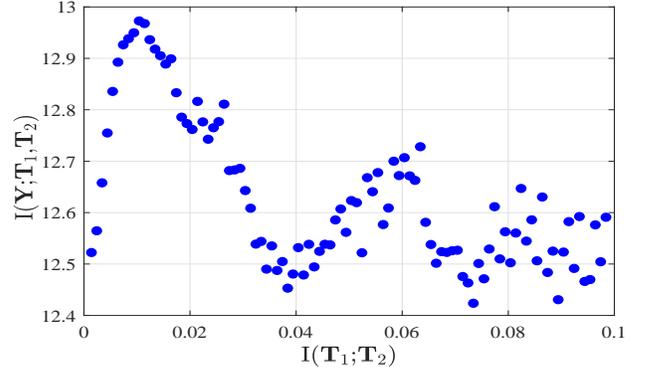}
			\caption{}
			\label{fig:plot4}
		\end{subfigure}
		\vspace{-0.2cm}
		\caption{The variation of compression levels constraints to ensure privacy vs total achievable prediction accuracy for multi-agent scenario. The individual affects of agent-$1$ and agent-$2$ privacy compression for collective prediction is shown in (a) and (b), respectively, and the joint effect is shown in (c). \vspace{-0.4cm}}
		\label{fig:simRes}
	\end{figure*}
	
	Using the iterative procedure outlined in Theorem\,\ref{thm:gaussianBottle} with suitable initial values for $\mathbf{\Sigma}_{\bm{\zeta}_{1}}, \mathbf{\Sigma}_{\bm{\zeta}_{2}}, \mathbf{A}$ and $\mathbf{B}$, we obtain the convergent solution for the DPFL. Upon varying the Lagrange parameters $(\beta, \lambda, \gamma)$, we obtain the variation of privacy inducing compression levels $I(\mathbf{X};\mathbf{T}_{1}), I(\mathbf{X};\mathbf{T}_{2})$ for agents\,${1},{2}$, respectively, with the collective prediction requirement $I(\mathbf{T}_{1}, \mathbf{T}_{2};\mathbf{Y})$ from the multi-agent setup. The numerical values of the associated information terms for $N_{\mathbf{X}} = 10, N_{\mathbf{Y}} = 10$ are shown in Figure\,\ref{fig:simRes}. We observe that, by relaxing the compression constraints on the agents (to ensure privacy) by increasing the $I(\mathbf{X};\mathbf{T}_{1}), I(\mathbf{X};\mathbf{T}_{2})$, the collective prediction accuracy $I(\mathbf{T}_{1}, \mathbf{T}_{2};\mathbf{Y})$ increases. The joint effect of the privacy constraint on the net achievable prediction is shown in Figure\,\ref{fig:plot3}. We observe the similar trend with relaxation of any agent's privacy constraint improves the multi-agent collective prediction accuracy.
	
	The role of possible correlation across the agents is shown in Figure\,\ref{fig:plot4}. We see that enabling small correlation across agents (i.e., $I({\bf T}_{1};{\bf T}_{2})>0$) helps in achieving higher collective prediction information. However, we see that excess correlation harms the predictive information. Intuitively, if the correlation is high across the agents, then for a given compression constraint, repeated compressed information is present across the agents and hence limits the collective prediction information. But, as observed, a small correlation helps because providing the agents with perfectly uncorrelated compressed information may not always be feasible and could adversely affect the prediction information.
	\section{Conclusion}
	\label{sec:concl}
	In this paper, we have considered the scenario of DL/FL, and we have addressed the problem of assuring privacy in a two-agent system while maximizing the prediction accuracy. We have formulated the problem for two-agent architecture as an information-theoretic approach. Assuming discrete alphabet, we have derived the stochastic scheme needed to be performed by the source to ensure privacy. The stochastic mappings are determined using a Blahut-Arimoto type of algorithm by alternating over self-consistent equations. Taking advantage of the problem formulation, we have proved that the provided algorithm converges to a stationary point. In addition, we have also extended our analysis to continuous distributions by assuming that the source's probability density function is Gaussian random variable and we have provided closed-form expressions for the alternating equations. As the presented framework shows promising results in assuring privacy in FL, a future direction to this approach is to ensure its scalability under reasonable complexity assumption.  
	
	\nocite{tishby2000information}
	
	\appendices
	
	\section{Iterative Procedure}
	\label{appn:iterProc}
	The iterative procedure using the self-consistent equations in Theorem\,\ref{thm:generalmultiAgent} at the $t$-th iteration is written as:
	\begin{align}
		&p^{(t+1)}(\mathbf{T}_{1}|\mathbf{X})=\frac{p^{(t)}(\mathbf{T}_{1})}{Z^{(t)}_1}\times\nonumber\\
		&\qquad\exp\left\{\frac{\gamma}{\beta}D_{KL}[p^{(t)}(\mathbf{T}_{2}\vert\mathbf{X})\vert\vert p^{(t)}(\mathbf{T}_{2}\vert\mathbf{T}_{1})] \right.\nonumber\\
		&\qquad \left. - \frac{1}{\beta}\mathbb{E}^{(t)}_{\mathbf{T}_{2}\vert\mathbf{X}}D_{KL}[p(\mathbf{Y}\vert\mathbf{X})\vert\vert p^{(t)}(\mathbf{Y}\vert\mathbf{T}_{1},\mathbf{T}_{2})]\right\},\label{eqn:SCeqn1}\\
		&p^{(t+1)}(\mathbf{T}_{2}|\mathbf{X})=\frac{p^{(t)}(\mathbf{T}_{2})}{Z^{(t)}_2}\times\nonumber\\
		&\quad\exp\left\{\frac{\gamma}{\lambda}D_{KL}[p^{(t+1)}(\mathbf{T}_{1}\vert\mathbf{X})\vert\vert p^{(t)}(\mathbf{T}_{1}\vert\mathbf{T}_{2})] \right.\nonumber\\
		&\quad \left. - \frac{1}{\lambda}\mathbb{E}^{(t+1)}_{\mathbf{T}_{1}\vert\mathbf{X}}D_{KL}[p(\mathbf{Y}\vert\mathbf{X})\vert\vert p^{(t)}(\mathbf{Y}\vert\mathbf{T}_{2},\mathbf{T}_{1})]\right\},\label{eqn:SCeqn2}\\
		&p^{(t+1)}(\mathbf{T}_{2}\vert\mathbf{T}_{1}) = \sum\limits_{\mathbf{X}}p^{(t+1)}(\mathbf{T}_{2}\vert\mathbf{X})p^{(t+1)}(\mathbf{X}\vert\mathbf{T}_{1}),\label{eqn:SCeqn3}\\
		&p^{(t+1)}(\mathbf{T}_{1}\vert\mathbf{T}_{2}) = \sum\limits_{\mathbf{X}}p^{(t+1)}(\mathbf{T}_{1}\vert\mathbf{X})p^{(t+1)}(\mathbf{X}\vert\mathbf{T}_{2}),\label{eqn:SCeqn4}\\
		&p^{(t+1)}(\mathbf{Y}\vert\mathbf{T}_{1},\mathbf{T}_{2}) = \sum\limits_{\mathbf{X}}p(\mathbf{Y}\vert\mathbf{X})p^{(t+1)}(\mathbf{X}\vert\mathbf{T}_{1},\mathbf{T}_{2}).
		\label{eqn:SCeqn5}
	\end{align}
	The equation (\ref{eqn:SCeqn5}) can be computed using the Bayes' rule and the conditional independence of $\mathbf{T}_1\perp\mathbf{T}_{2}\vert\mathbf{X}$ as follows.
	\begin{equation*}
		p(\mathbf{X}\vert\mathbf{T}_{1},\mathbf{T}_{2}) = \frac{p(\mathbf{X})p(\mathbf{T}_{1}\vert\mathbf{X})p(\mathbf{T}_{2}\vert\mathbf{X})}{\sum\limits_{\mathbf{\tilde{X}}}p(\mathbf{\tilde{X}})p(\mathbf{T}_{1}\vert\mathbf{\tilde{X}})p(\mathbf{T}_{2}\vert\mathbf{\tilde{X}})}.
	\end{equation*}
	
	\section{Definitions for Theorem\,\ref{thm:gaussianBottle}}
	\label{appn:paramDefGaussian}
	The parameters used in Theorem\,\ref{thm:gaussianBottle} are defined in this Section. For the jointly Gaussian variables $\mathbf{X}$ and $\mathbf{Y}$, we denote $\bm{\Theta} = \mathbf{\Sigma}_{\mathbf{Y}\,\mathbf{X}}\mathbf{\Sigma}_{X}^{-1}$. The rest of the variables are defined as follows.
	\begin{align*}
		&\bm{\Xi}_{1} = \mathbf{B}\mathbf{\Sigma}_{\mathbf{X}}\mathbf{A}^{T}\mathbf{\Sigma}_{\mathbf{T}_{1}}^{-1}\\
		&\bm{\Xi}_{2} = \mathbf{A}\mathbf{\Sigma}_{\mathbf{X}}\mathbf{B}^{T}\mathbf{\Sigma}_{\mathbf{T}_{2}}^{-1},\\
		&\bm{\Psi}_{1} = \bm{\Theta}(\mathbf{\Sigma}_{X}\mathbf{A}^{T}\mathbf{\Sigma}_{\mathbf{T}_{1}\vert\mathbf{T}_{2}}^{-1} - \mathbf{\Sigma}_{X}\mathbf{B}^{T}\mathbf{\Sigma}_{\mathbf{T}_{2}\vert\mathbf{T}_{1}}^{-1}\mathbf{B}\mathbf{\Sigma}_{X}\mathbf{A}^{T}\mathbf{\Sigma}_{\mathbf{T}_{1}}^{-1}),\\
		&\bm{\Psi}_{2} = \bm{\Theta}(\mathbf{\Sigma}_{X}\mathbf{B}^{T}\mathbf{\Sigma}_{\mathbf{T}_{2}\vert\mathbf{T}_{1}}^{-1} - \mathbf{\Sigma}_{X}\mathbf{A}^{T}\mathbf{\Sigma}_{\mathbf{T}_{1}\vert\mathbf{T}_{2}}^{-1}\mathbf{A}\mathbf{\Sigma}_{X}\mathbf{B}^{T}\mathbf{\Sigma}_{\mathbf{T}_{2}}^{-1}).
	\end{align*}
	Also, using the Schur's complement definition, we derive the following.
	\begin{align*}
		&\mathbf{\Sigma}_{\mathbf{X}\vert\mathbf{T}_{1},\mathbf{T}_{2}} = \mathbf{\Sigma}_{\mathbf{X}} - \mathbf{\Sigma}_{\mathbf{X}}\mathbf{A}^{T}\mathbf{\Sigma}_{\mathbf{T}_{1}\vert\mathbf{T}_{2}}^{-1}\mathbf{A}\mathbf{\Sigma}_{\mathbf{X}} \\
		&\hspace*{50pt}+ \mathbf{\Sigma}_{X}\mathbf{A}^{T}\mathbf{\Sigma}_{\mathbf{T}_{1}\vert\mathbf{T}_{2}}^{-1}\mathbf{A}\mathbf{\Sigma}_{X}\mathbf{B}^{T}\mathbf{\Sigma}_{\mathbf{T}_{2}}^{-1}\mathbf{B}\mathbf{\Sigma}_{X} \\
		&\hspace*{50pt}+ \mathbf{\Sigma}_{X}\mathbf{B}^{T}\mathbf{\Sigma}_{\mathbf{T}_{2}\vert\mathbf{T}_{1}}^{-1}\mathbf{B}\mathbf{\Sigma}_{X}\mathbf{A}^{T}\mathbf{\Sigma}_{\mathbf{T}_{1}}^{-1}\mathbf{A}\mathbf{\Sigma}_{X} \\
		&\hspace*{50pt} - \mathbf{\Sigma}_{\mathbf{X}}\mathbf{B}^{T}\mathbf{\Sigma}_{\mathbf{T}_{2}\vert\mathbf{T}_{1}}^{-1}\mathbf{B}\mathbf{\Sigma}_{\mathbf{X}}.
	\end{align*}
	The required covariance matrix $\mathbf{\Sigma}_{\mathbf{Y}\vert\mathbf{T}_{1},\mathbf{T}_{2}}$ can now be written as 
	\begin{equation*}
		\mathbf{\Sigma}_{\mathbf{Y}\vert\mathbf{T}_{1},\mathbf{T}_{2}} = \bm{\Theta}\mathbf{\Sigma}_{\mathbf{X}\vert\mathbf{T}_{1},\mathbf{T}_{2}}\bm{\Theta}^{T} + \mathbf{\Sigma}_{\mathbf{Y}\vert\mathbf{X}}.
	\end{equation*}
	Furthermore, using matrix inversion lemma, we can write that
	\begin{eqnarray*}
		\mathbf{\Sigma}_{\mathbf{X}\vert\mathbf{T}_{1}} &=& \mathbf{\Sigma}_{X} - \mathbf{\Sigma}_{X}\mathbf{A}^{T}(\mathbf{A\mathbf{\Sigma}_{X}\mathbf{A}^{T} + \mathbf{\Sigma}_{\bm{\zeta}_{1}}})^{-1}\mathbf{A}\mathbf{\Sigma}_{X}\\
		&=& (\mathbf{\Sigma}_{X}^{-1} + \mathbf{A}^{T}\mathbf{\Sigma}_{\bm{\zeta}_{1}}^{-1}\mathbf{A})^{-1},
	\end{eqnarray*}
	and then we express, $\mathbf{\Sigma}_{\mathbf{T}_{2}\vert\mathbf{T}_{1}} = \mathbf{B}\mathbf{\Sigma}_{\mathbf{X}\vert\mathbf{T}_{1}}\mathbf{B}^{T} + \mathbf{\Sigma}_{\bm{\zeta}_{2}}$. Using symmetry, we can also write the definition for covariance matrix $\mathbf{\Sigma}_{\mathbf{T}_{1}\vert\mathbf{T}_{2}}$.
	\bibliography{multiAgents}
	\bibliographystyle{IEEEtran}
\end{document}